\documentclass[epj,referee]{svjour}
\usepackage{epsfig}
\usepackage{graphicx}
\usepackage{amssymb}

\begin{document}
\title{Buckling of swelling gels}
\author{Thierry Mora\thanks{\emph{Present address: LPTMS, Universit\'e Paris-Sud, 
Centre scientifique d'Orsay, 15 rue Georges Cl\'emenceau, 91405 Orsay cedex, France.} }
\and Arezki Boudaoud
%
}                     
\offprints{A. B., boudaoud@lps.ens.fr.}          
\institute{Laboratoire de Physique Statistique de l'\'Ecole Normale Sup\'erieure, 24, rue Lhomond, 75231 Paris Cedex 05, France.}

\date{Received: date / Revised version: date}

\abstract{
The patterns arising from the differential swelling of gels are investigated experimentally and theoretically as a model for the differential growth of living tissues. Two geometries are considered: a thin strip of soft gel clamped to a stiff gel, and a thin corona of soft gel clamped to a disk of stiff gel. When the structure is immersed in water, the soft gel swells and bends out of plane leading to a wavy periodic pattern which wavelength is measured. The linear stability of the flat state is studied in the framework of linear elasticity using the equations for thin plates. The flat state is shown to become unstable to oscillations above a critical swelling rate and the computed wavelengths are in quantitative agreement with the experiment.
} 
\maketitle

\section{Introduction}
\label{intro}
Living organisms are full of fascinating complex patterns. One might wonder about the physical mechanisms at stake as well as their relevance. Although a tissue is obviously an elastic solid, the role of mechanical stresses in morphogenesis was not recognised until recently. On the one hand they were shown to be important in phyllotaxis (the arrangement of leaves in plants)~\cite{steele,shipman}, in the wrinkling of leaves~\cite{sharon,audoly03}, in the selection of cell sizes~\cite{boudaoud03a}, as well as in the development of embryos~\cite{brouzes}. On the other hand, a theoretical framework was introduced to study the instabilities occuring in the growth of elastic bodies~\cite{benamar05}. 

At first sight, it is difficult to find physical systems allowing the investigation of growth. The tearing of plastic sheets as in~\cite{sharon} gives little control over the growth rate. However some polymeric gels can undergo huge volume changes when submitted to external stimuli such as variations in temperature, pH, osmotic pressure \cite{tanaka78,tanaka80}, electric field \cite{tanaka82} or light \cite{juodkazis}. Strictly speaking, such a gel does not grow but swells by absorbing water while its elastic modulus decreases.  A number of studies (see~\cite{boudaoud03b} for a review) have been devoted to the instabilities of swelling or deswelling gels, such as the folding of the surface of swollen gels clamped to hard substrates and the subsequent formation of a network of cusp lines~\cite{tanaka87,hwa,onuki89,suematsu90,htanaka92}.

In this article we are concerned with the instabilities occuring in the differential swelling of gels. Our motivation is to design physical counterparts to growing tissues. We use thin gel plates made by assembling two gels with different elastic and swelling properties and we investigate the resulting patterns experimentally and theoretically. First, in section~\ref{experiment}, we describe the experimental procedure and give our first observations. In section~\ref{theory}, we introduce the theoretical framework and the linear stability analysis. In section~\ref{results}, we compare quantitatively the experimental and theoretical results. We eventually give a discussion and some perspectives in section~\ref{conclusion}.

\section{The experiment}
\label{experiment}
Our goal was to design an experiment to study the differential swelling of gels. We chose to assemble two thin flat gels having different elastic and swelling properties. Polyacrylamide gels were suitable as they swell when immersed in water whereas their swelling and elastic properties can be tuned independently.

We  prepared our gels as in~\cite{tanaka78,tanaka80,li}. A mixture of acrylamide (AA) and N,N'-methylenebisacrylamide (BISAA) is dissolved with sodium acrylate (SA) in distilled water. The polymerization is initiated by ammonium persulfate (PA) and is catalysed with N,N,N',N'-tetramethylenediamine (TEMED) (0.3\% in volume).  The composition of the gels used mostly is given in Tab.~\ref{typesgel}.  In these conditions, gelation generally occurs within one minute after the addition of the catalyst.

The characteristics of the gel can be tuned by varying the concentrations of the components. The more concentrated (and, for a same concentration, the more concentrated in BISAA) the solution, the stiffer the gel. Likewise, the swelling rate can be increased by adding sodium acrylate. For the purpose of the experiment we have prepared two distinct types of gel: (I) a soft and swelling gel; (II) a stiff and nonswelling gel. The elastic and swelling properties of these gels have been measured and are reported in Tab.~\ref{typesgel}.

\begin{table}[h]
\begin{center}
\begin{tabular}{l c c}
{} & I & II\\
\noalign{\smallskip}\hline\noalign{\smallskip}
$[$AA+BISAA$]$ & 720 & 2880\\
BISAA:AA ratio & 1:37.5 & 1:19\\
$[$SA$]$ & 46-183 & 0\\
Swelling rate & 50\%-80\% & 6\%\\
$E$ (Pa) & $5.0\cdot 10^3$ & $3.2\cdot 10^5$ \\
\end{tabular}
\end{center}
\caption{Composition and properties of the gels used in the experiment. Concentrations are given in mmol.L${}^{-1}$. The swelling rate corresponds to the increase in linear dimensions. $E$ is the elastic modulus of the gel.}
\label{typesgel}
\end{table}

\begin{figure}
\begin{center}
\epsfig{file=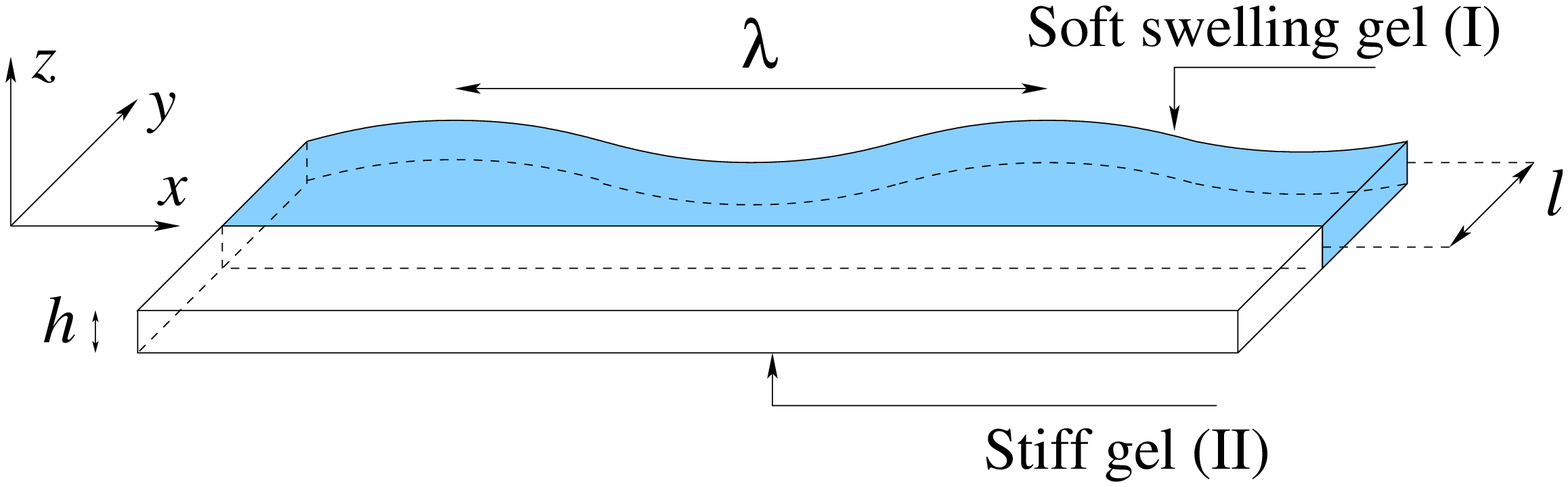,width=.95\columnwidth}
\epsfig{file=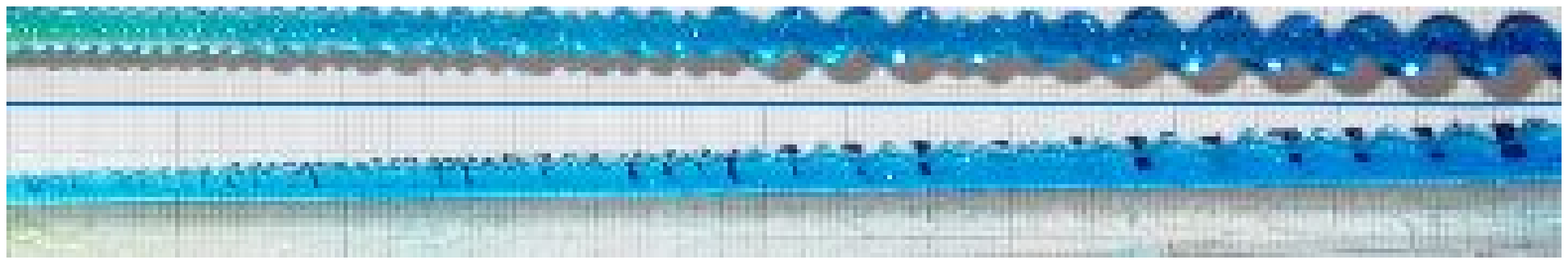,width=.95\columnwidth}
\caption{The strip geometry. Schematic, top and side pictures. A strip of soft swelling gel (I) is chemically clamped to a strip of stiff nonswelling gel (II). When immersed in water, part (I) swells while part (II) does not. An off-plane instability appears on (I) in the $x$ direction if the width $l$ is large enough (there are regularly spaced cusps otherwise). Its wavelength $\lambda$ increases monotonically with the width $l$ as seen in pictures.}
\label{schema}
\end{center}
\end{figure}

\begin{figure}
\begin{center}
\epsfig{file=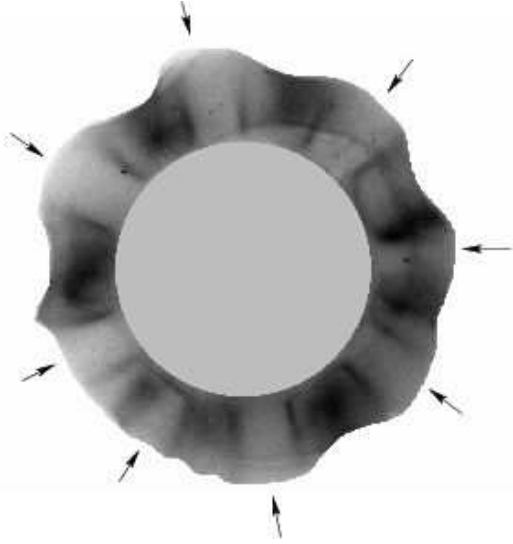,width=.95\columnwidth}
\caption{The corona geometry. A corona (inner radius $r_i$ and outer radius $r_o$) of soft swelling gel (I) is clamped to a disk of stiff nonswelling gel (II). When swelling, the corona becomes unstable and goes off the inital plane. The wavenumber (eight on this picture) increases monotonically with the aspect ratio $r_i/r_o$.}
\label{schemarondelle}
\end{center}
\end{figure}

The gelation process is performed in a thin cell composed of two glass plates separated by rubber spacers of constant thickness $h$ (1 to 5 mm). The cell is set vertically. The spacers are used as masks shaped according to two different geometries: 
\begin{itemize}
	\item The so-called \emph{strip geometry} consists of two thin strips of gel of respective compositions I and II, clamped by their edges (see Fig.~\ref{schema}). To obtain this geometry, solution II was first poured into the cell and left for gelation. Then solution I was added to form a second layer of width $l$ (typically 1 cm) and length 20 cm. In the process of gelation of the second layer, the two layers become chemically clamped to each other.
	\item The \emph{corona geometry} is the axisymmetric counterpart of the strip geometry. A disk (radius $r_i$ in the range 2--5 cm) of gel II is clamped to a corona (inner radius $r_i$, outer radius $r_o$ in the range 2--5 cm as well) of gel I (see Fig.~\ref{schemarondelle}). We used circular masks of various radii to obtain these shapes: first the disk is made, then the cell is opened, the mask is replaced and the corona is moulded after closing back the cell.
\end{itemize}

The obtained structure is removed from the cell and immersed in water for several hours until a stationary state is reached and the swelling process is complete. In both cases, the sample undergoes a mechanical buckling instability such that the soft gel is no longer flat (except for narrow strips -- see below). This  results in a wavy pattern  (Figs.~\ref{schema}--\ref{schemarondelle}) with a well-defined wavelength. 

In the strip geometry, the wavelength increases with the width of the strip $l$. We also performed some experiments with different swelling ratios and found similar results. For a small width $l$ no buckling occurred but we observed instead a fine pattern with regularly spaced cusps similar to those studied in~\cite{tanaka87}. In the corona geometry, the pattern is periodic and is characterised by a wavenumber (defined as the number of complete wavelengths) which  increases with the aspect ratio $r_o/r_i$. A more quantitative description is delayed to section~\ref{results}. 

\section{Theoretical setting}\label{theory}

We now study the patterns using a linear stability analysis of the equations of elasticity. We consider only the soft swelling part (I) of the structure, the other part (II) being considered as static as its swelling rate is small and it is very stiff. Since the gel sample is thin and flat before the instability, we can use the Foppl - von K\'arm\'an equations for thin elastic plates~\cite{landau}. Material points are parametrized by their initial planar cartesian coordinates $x$ and $y$. A deformation is defined by the displacement field $(u_x(x,y),u_y(x,y),\zeta(x,y))$; $u_x$ and $u_y$ are the in-plane displacements along the $x$ and $y$ axes respectively whereas $\zeta$ is the transverse (off-plane) displacement. 
We use the framework of linear elasticity which will prove sufficient for the interpretation of the results (this restriction is discussed in the conclusion), so that the in-plane stress tensor $\sigma_{\alpha\beta}$ ($\alpha,\beta=x,y$)  depends linearly on the deformation tensor $u_{\alpha\beta}$~\cite{landau}:
\begin{equation}
u_{\alpha \beta}=\frac{1}{2}\left(\frac{\partial u_\alpha}{\partial x_\beta}+\frac{\partial u_\beta}{\partial x_\alpha}+\frac{\partial \zeta}{\partial x_\alpha}\frac{\partial \zeta}{\partial x_\beta}\right),
\end{equation}
\begin{eqnarray}
\sigma_{xx}&=&\frac{E}{1-\sigma^2}\left(u_{xx}+\sigma u_{yy}\right),\\
\sigma_{yy}&=&\frac{E}{1-\sigma^2}\left(u_{yy}+\sigma u_{xx}\right),\\
\sigma_{xy}&=&\frac{E}{1+\sigma}u_{xy},
\end{eqnarray}
where $\sigma=1/2$ is the Poisson ratio of the gel (these gels are almost incompressible). 
The F\"oppl - von K\'arm\'an equations for equilibrium~\cite{landau} read
\begin{eqnarray}\label{elast}
D\Delta^2\zeta-h\frac{\partial}{\partial x_\beta}\left(\sigma_{\alpha\beta}\frac{\partial \zeta}{\partial x_\alpha}\right)=0, \\
\frac{\partial \sigma_{\alpha\beta}}{\partial x_\beta}=0, \label{stress}
\end{eqnarray}
for out of plane bending and in-plane stretching respectively. The bending stiffness is
\begin{equation}
\quad\textrm{where}\quad D=\frac{Eh^3}{12(1-\sigma^2)}.
\end{equation}
The corresponding (linearized) boundary conditions are given in the next two subsections.

\subsection{The strip geometry}

We now specialise to the strip geometry. The reader should refer to Fig. \ref{schema} for the notations. From this point onwards we use $l$ as unit of length and $D/(hl^2)$ as unit of stress. The constraint imposed by II results in the compression of the swollen strip in the $x$ direction. We study the linear stability of the flat solution such that $\zeta=0$. Using the translational invariance of the system we look for in-plane deformations in the form $u_x=-kx$ ($k>0$), $u_y=f(y)$. The boundary conditions $u_y(0,x)=0$ at the clamped edge and $\sigma_{xx}(x,1)=0$ at the free edge, along with (\ref{stress}), impose $u_y=0$. Then the equilibrium equations reduce to
\begin{equation}\label{elaststrip}
\Delta^2\zeta-k\frac{\partial^2\zeta}{{\partial x}^2}=0.
\end{equation}

We look for periodic solutions in the form 
\begin{equation}
\zeta(x,y)=\xi(y)\cos qx.
\end{equation}
Using~(\ref{elaststrip}), $\xi(y)$ appears to be the solution of a fourth-order linear differential equation with constant coefficients. Its solutions can be written in the form
\begin{equation}
\xi(y)=A \exp(imy)+ B \exp(-imy) + C\exp(ny)+D\exp(-ny),
\end{equation}
with
\begin{equation}\label{condi}
k=\frac{{\left(m^2+q^2\right)}^{2}}{q^2}\quad\textrm{and}\quad n^2-m^2=2q^2.
\end{equation}

The boundary equations on the clamped ($y=0$) and free ($y=1$) edges, read \cite{landau}:
\begin{displaymath}\label{boundstrip}
\begin{array}{ll}
\xi(0)=0, & \quad\xi''(1)-\sigma q^2\xi(1)=0, \\
\xi'(0)=0, & \quad-\xi^{(3)}(1)+(2-\sigma)q^2\xi'(1)=0.
\end{array}
\end{displaymath}
These conditions can be viewed as a system of 4 linear equations with the four unknowns $A, B, C, D$. A non-zero solution $\zeta$ exists if and only if  the determinant is zero, which occurs for a certain $n(q)$ (we recall that $m$ can be expressed as a function of $m$ and $q$ from Eq.~(\ref{condi})).

\begin{figure}
\begin{center}
\epsfig{file=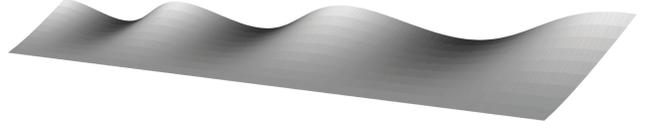,width=.95\columnwidth}
\caption{\label{simulstrip} Most unstable mode for an infinite flat strip of gel longitudinally compressed on one edge, as given by the solution to (\ref{elaststrip}).}
\end{center}
\end{figure}

We now look for the most unstable wavenumber $q_c$, \textit{i. e.} the one for which a non-zero solution exists with the lower value of the length increase $k=k_c$. Thus we implicitly assume that $k$ is near its threshold value $k_c$. This is another limitation which is discussed in the conclusion.
The quantity $k={\left(n(q)^2+q^2\right )}^2/q^2$ is then minimized with respect to $q$. Eventually the wavelength reads  (in dimensional units)
\begin{equation}\label{stripth}
\lambda=\frac{2\pi}{q_c}=3.256\, l.
\end{equation}
A 3-dimensional representation of the corresponding solution $\zeta(x,y)$  of (\ref{elaststrip}) is shown in Fig.~\ref{simulstrip}.

\subsection{The corona geometry}
We now proceed to the same analysis for the corona geometry. The swollen corona ($r_i<r<r_o$) is supposed to be radially pulled at the inner edge so that its inner radius shrinks to fit the radius of the unswollen disk. Note that $r_i$ and $r_o$ are the dimensions of the corona \emph{if it had swollen unconstrained}. Although neither of these radii can be measured directly, their ratio $r_i/r_o$ is the same as before swelling. From this point onwards we use the outer radius $r_o$ as unit of length and $D/(h r_o^2)$ as unit of stress. In these reduced units, $r_i$ is the aspect ratio of the corona before the swelling.

We first need the stress field in the flat corona as resulting from the tension at the inner radius. We look for a solution of (\ref{stress})  in the form:
\begin{equation}
u_r=ar+\frac{b}{r},\quad u_\theta=0.
\end{equation}
$r$ and $\theta$ are the standard polar coordinates. The boundary conditions are $u_r(r=r_i)=-\beta$ (displacement to fit the unswollen gel) at the inner edge and $\sigma_{rr}(r=1)=0$ (stress free) at  the outer edge. The problem has now one single degree of freedom $\beta$ . We obtain a stress tensor of the form:
\begin{equation}
\sigma_{rr}=-\alpha\left(1-\frac{1}{r^2}\right),\quad\sigma_{\theta\theta}=-\alpha\left(1+\frac{1}{r^2}\right).
\end{equation}
The number $\alpha$ being proportional to $\beta$ is chosen as the swelling control parameter.
The balance of moments (\ref{elast}) reduces to
\begin{equation}\label{elastcorona}
{\Delta}^2 \zeta + \alpha\Delta \zeta + \frac{\alpha}{r^2}\left(-\frac{{\partial}^2\zeta}{\partial r^2}+\frac{3}{r}\frac{\partial \zeta}{\partial r}+\frac{1}{r^2}\frac{{\partial}^2\zeta}{\partial {\theta}^{2}}\right)=0,
\end{equation}
\begin{equation}
\textrm{with}\quad\Delta\zeta=\frac{{\partial}^2\zeta}{\partial r^2}+\frac{1}{r^2}\frac{{\partial}^2\zeta}{\partial {\theta}^2}+\frac{1}{r}\frac{\partial \zeta}{\partial r}.
\end{equation}
At the clamped edge ($r=r_i$), the boundary conditions read 
\begin{eqnarray}
\xi=0,\nonumber \\ \frac{\partial\xi}{\partial r}=0,\label{bound1}
\end{eqnarray}
whereas at the free edge ($r=1$),
\begin{eqnarray}
-\frac{\partial}{\partial r}\Delta\zeta + (1-\sigma )\frac{1}{r^3}\left(\frac{{\partial}^2\zeta}{\partial{\theta}^2}-r\frac{{\partial}^3\zeta}{\partial r \partial{\theta}^2}\right)&=&0,\nonumber\\
\Delta \zeta + (\sigma -1)\frac{1}{r^2}\left(\frac{{\partial}^2\zeta}{\partial{\theta}^2}+r\frac{\partial \zeta}{\partial r}\right)&=&0.\label{bound2}
\end{eqnarray}

We look for periodic solutions to~(\ref{elastcorona}) in the form
\begin{equation}
\zeta(r,\theta)=\xi(r)\cos m\theta,
\end{equation}
where the wavenumber $m$ is an integer. We find a fourth-order linear equation in $\xi(r)$. 

\begin{figure}
\begin{center}
\includegraphics[width=0.95 \columnwidth]{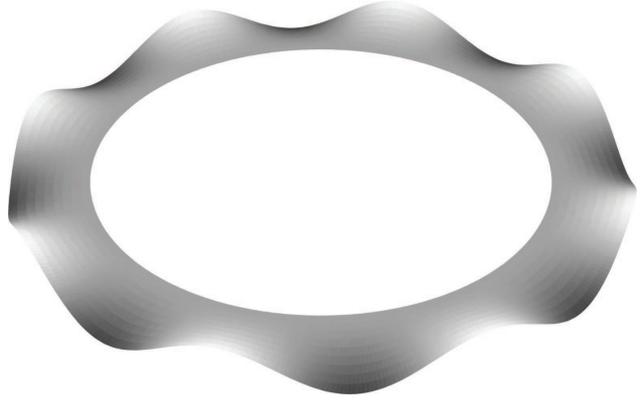}
\caption{\label{simulcorona} Most unstable mode for a flat corona of gel with radial tension at the inner radius. The aspect ratio (inner radius/outer radius) is $0.74$. The selected wavenumber is $m=8$.}
\end{center}
\end{figure}

We first compute a basis of the 2-dimensional space formed by the solutions satisfying the conditions at the inner boundary (\ref{bound1}). To proceed, we solve the differential equation with initial conditions:
\begin{displaymath}
(\xi(r_i),\xi'(r_i),\xi''(r_i),\xi'''(r_i))=\left\{\begin{array}{l}
(0,0,1,0)\\
(0,0,0,1)\end{array}\right.,
\end{displaymath}
using a Runge-Kutta algorithm, so that we find the basis $\xi_1,\xi_2$.
A non-zero linear combination of $\xi_1$ and $\xi_2$ verifying the conditions at the outer boundary (\ref{bound2}) exists only for a certain $\alpha(m,r_i)$, which we compute numerically. Using the same argument as in the strip geometry, we choose $m(r_i)$ for which $\alpha(m,r_i)$ is minimum. Thus we obtain the wavenumber $m$ as a function of the aspect ratio $r_i$. Fig. \ref{simulcorona} shows the solution $\xi(r,\theta)$ of (\ref{elastcorona}) with the aspect ratio $r_i/r_o=0.74$ for which $m=8$.

We also tested the $r_i\to 1$ limit of our numerical calculation. In this limit, the corona geometry should reduce to the strip geometry, as the curvature of the interface between the two gels vanishes. The wavelength reads 
\begin{equation}
\frac{\lambda}{l}\approx\frac{\pi}{m}\frac{1+r_i}{1-r_i}.
\end{equation}
For $r_i=0.9999$, we obtained indeed $\lambda/l=3.255$ in agreement with (\ref{stripth}).

\section{Results}\label{results}

\subsection{The strip geometry}

\begin{figure}
\begin{center}
\epsfig{file=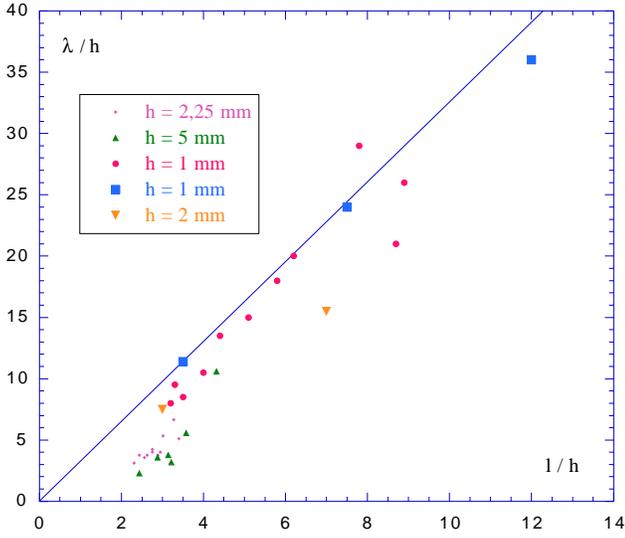,width=.95\columnwidth}
\caption{\label{stripplot} Instability wavelength $\lambda$ as a function of the width $l$ of the swollen strip. The measurements are normalised by the thickness $h$ of the gel. Off-plane instability is observed for $l>l_c\sim 2h$. For $l<l_c$, cusped patterns appear on the surface of the swollen gel. The line corresponds to the theoretical result $\lambda=3.256\,l$ (Eq.~\ref{stripth} -- valid for $l\gg h$).}
\end{center}
\end{figure}

The two geometrical parameters are the width $l$ of the swollen strip, and the thickness $h$ of the gel before swelling.  We plotted our observations on Fig. \ref{stripplot}, along with the analytical result (\ref{stripth}). The gel goes off the plane for $l>l_c\sim 2h$. The wavelength of the instability increases with $l$, and is approximately linear with $l$ in the limit of small thickness $h\ll l$, in agreement with the theoretical result.
For $l<l_c$, instability patterns are observed on the surface of the swollen gel. They consist in regularly spaced cusps similar to those observed in~\cite{tanaka87}.

\subsection{The corona geometry}

\begin{figure}
\begin{center}
\epsfig{file=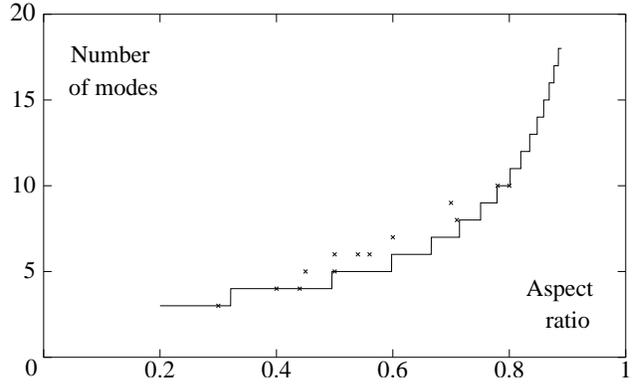,width=.95\columnwidth}
\caption{\label{coronaplot} Wavenumber $m$ of the instability as a function of the aspect ratio  (inner radius $r_i$ / outer radius $r_o$) before swelling. Crosses: experimental results. Line: theoretical result.}
\end{center}
\end{figure}

In the corona geometry experiment, the thickness $h$ was fixed to 1 mm, and the inner and outer radii were varied from 20 mm to 50 mm. Thus the theory of thin plates remains valid, which allows a direct comparison between experimental and analytical results (Fig. \ref{coronaplot}). The wavenumber $m$ increases with the aspect ratio $r_i/r_o$, in quantitative agreement with the theoretical predictions. 

\section{Conclusion}\label{conclusion}
To summarise, we showed that the swelling of thin soft gel plates clamped to a stiff gel leads to a buckling instability. A linear stability analysis yields a prediction for the pattern wavelengths in quantitative agreement with the experiment. These wavelengths are mainly determined by the in-plane geometry of the thin gel. Our analysis is restricted by two main limitations.

 On the one hand, we used the most unstable modes to predict the wavelengths at the instability threshold whereas the experimental conditions are far above this threshold. However standard weakly nonlinear analysis generally yields almost the same wavelengths~\cite{manneville}. On the other hand, we used the framework of linear elasticity while the swelling rate is large and the deformations are finite. Moreover the flat base state is anisotropic: for instance, in the strip geometry, the gel has swollen differently along the two in-plane direction, so that the elastic modulus is not strictly the same in these two directions. The small discrepancies between the experimental and theoretical wavelengths might be ascribed to either of these two limitations. 
 
 A suggestion for future research stems from the second limitation. One might use the framework developed in~\cite{benamar05} to build a more precise theory for the present experiment. On the experimental side, we have shown how to design physical counterparts to growing tissues. Other geometries are currently investigated and will be the subject of forthcoming publications.

\begin{acknowledgement}
We are grateful to Laurent Quartier for his generous help with the experiment. This study was partially supported by the Minist\`ere de la Recherche--ACI Jeunes Chercheurs and by the European Community--New and Emerging Science and Technology programs. LPS is UMR 8550 of CNRS and is associated with Universit\'e Pierre et Marie Curie (Paris VI) and Universit\'e Denis Diderot (Paris VII).
\end{acknowledgement}

\bibliographystyle{unsrt}
\bibliography{article}

\end{document}